# Hyper-Selective Plasmonic Color Filters

Dagny Fleischman[†], Luke A. Sweatlock[†*], Hirotaka Murakami[‡], Harry Atwater[†]

[†]California Institute of Technology, 1200 E California Blvd, Pasadena CA 91125

[*]Northrop Grumman NG Next, One Space Park, Redondo Beach CA 90250

[‡]Sony Corporation 1-7-1 Konan, Minato-ku, Tokyo, 108-0075 Japan

**ABSTRACT:** The subwavelength mode volumes of plasmonic filters are well matched to the small size of state-of-the-art active pixels (~ 1 μm) in CMOS image sensor arrays used in portable electronic devices. Typical plasmonic filters exhibit broad (> 100 nm) transmission bandwidths. Dramatically reducing the peak width of filter transmission spectra would allow for the realization of CMOS hyperspectral imaging arrays, which demand the FWHM of transmission peaks to be less than 30 nm. We find that the design of 5 layer metal-insulator-metal-insulator-metal structures gives rise to multi-mode interference phenomena that suppresses spurious transmission features gives rise to a single narrow transmission band with FWHM as small as 17 nm. The transmission peaks of these multilayer slot-mode plasmonic filters (MSPFs) can be systematically varied throughout the visible and near infrared spectrum, so the same basic structure can serve as a filter over a large range of wavelengths.

Periodic arrays of subwavelength holes or nanoslits in metal films enable efficient conversion of optical energy between incident photons and surface propagating two-dimensional charge density waves, surface plasmon polaritons (SPPs). Due to the permittivity discontinuity at metal-dielectric surfaces, SPPs have an in-plane momentum $k_{SPP}$ greater than that of light in free space $k_o$. Patterned metal surfaces including gratings, or arrays of holes or slits, allow the matching of momentum and thereby enable efficient conversion of light into SPPs via scattering. The strength of interaction between photons and SPPs can be tailored by changing geometric factors such as the shape of the scattering elements, and the symmetry and periodicity of the array as well as by selecting the permittivity of the constituent materials[1].

In particular, periodic arrays of subwavelength apertures passing through a metal film exhibit enhanced transmission exclusively at conditions corresponding to constructive mutual interference between incident light and SPPs traveling along the surface between adjacent slits. In the case that the metallic layer is thick enough to be substantially opaque to incident photons, the SPP mediated process is the dominant mode of transmission and the surface acts as a band-pass color transmission filters. Such aperture arrays have been the topic of substantial scientific interest due to these remarkable optical properties and their utility as a testbed for studying fundamental light-matter interactions in plasmonic systems[2,3]. Due to their simplicity, scalability, and durability, plasmonic slit gratings have become an attractive route for development of technological applications including ultra-compact filters suitable for cameras and displays[4,5].

The dispersion of plasmonic propagating modes can be further engineered using metal-clad slot waveguides, often realized as multilayer stacks with a metal-insulator-metal (MIM) configuration[6]. Such MIM stacks may support a multitude of polaritonic modes which lie either inside or outside the "light cone," that is, with in-plane momentum either greater or less than that of a photon with equal energy. This additional degree of freedom enables substantially more complex optical transmission filter spectra enabling narrow bandwidth suitable for multispectral and hyperspectral color filtering applications[7].

**Designing Plasmonic Color Filters**

Finite difference time domain methods (FDTD) were used to determine the transmission spectra of different filter structures. Figure 1 illustrates the different types of transmission filters and their spectral behavior. MIM have been used to make RGB color filters[7]. These structures can be optimized to have narrowband transmission, but as the structure is optimized to minimize FWHM of the transmission peak, the intensity of the next highest order mode increases. This trade-off can be lifted by introducing a second MIM mode into the structure that couples with the original MIM mode, leading to the suppression of the spurious transmission. The asymmetric nature of the coupled MIM modes plays a role in the suppression of the spurious transmission, as illustrated in Supplemental Figure 1(a). The multilayer slot-mode plasmonic filter (MSPF) investigated demonstrates a narrow transmission bandwidth and spurious peak suppression, as shown in Figure 1(b), and by changing the periodicity of the slits, this filter can be swept across the entire visible spectrum.

The MSPFs were optimized using parameters sweeps that considered both the thicknesses and optical indices of all the insulating and metallic layers, as well as the width and spacing of the milled slits. The initial values for the thicknesses of the metallic layers were determined by considering the skin and penetration depths of various metals. For a successful filter, the top and bottom metallic layers of the structure must be sufficiently thick to be opaque across the visible and near IR parts of the electromagnetic spectrum. Using Rakic data for Ag as an example, the 1/e penetration depth ($d_p$) of Ag was calculated to range from 12.9 nm – 16.8 nm across the visible spectrum. To prevent 98% of light from penetrating the structure, the top and bottom layers must be 4 times $d_p$, so 51.6 nm to 67.2 nm. Therefore 68 nm was used as the initial parameter sweep value when optimizing the system that utilized Ag.

Likewise, the starting point for the thickness of the insulating layers was approximated by considering the propagating modes guided laterally within the structure. Numerically determined dispersion curves derived from experimental optical constants of Ag and $SiO_2$ can be used to determine the available modes within an MIM[6]. For $SiO_2$ thicknesses of less than 100 nm, traditional photonic waveguide modes are cut off in an Ag/$SiO_2$/Ag system, so the waveguide only supports high-momentum surface plasmon modes. Therefore, the parameter sweeps used 100 nm as the upper value restriction for the $SiO_2$ thickness of each waveguide.

Iterating over the parameter sweep led to the final device structure, with alternating layers of Ag and $SiO_2$. Both $SiO_2$ layers were optimized to 70 nm, the top and bottom Ag layers are 70 nm and the spacer layer is 50 nm. The width of the slit is 50 nm for all filters and the slit periodicities investigated vary from 250 nm to 550.

The position of the transmission peak varies linearly with the periodicity of the slits and, as shown in Figure 2(a), peak position can be swept across the visible and near IR spectrum. Therefore, just by varying the inter-slit pitch, a series of MSPFs with the same layer materials and thicknesses can be used as a color filter across a wide range of the spectrum. The FWHM of the transmission spectra are about 20 nm on average with no peak exceeding 28 nm, as shown in Figure 2(b). Additionally, the overall transmission of the side-lobe peak does not exceed 11% of that of the primary peak in the visible portion of the spectrum, and does not exceed 25% of the primary peak intensity in all filters investigated.

**Analytical Analysis**

A series of FDTD simulations were executed by sweeping over the visible spectrum using a single frequency plane wave source. Complex vector field data was collected by finely meshed monitors capturing the EM behavior over the span of each of the FDTD simulations. This data contained the Cartesian space variations of the electric and magnetic field information over time, which evolved as the plane wave injected into the simulation interacted with the MSPF. This very large dataset can be compressed by taking a discrete-time Fourier transform at runtime to yield the field data in the frequency domain.

A single electric field component from the compressed data set of a single simulation is plotted in Figure 3(a). The spatial mapping of the electric field superimposed on the MSPF depicts multiple modes that are active in the filter structure. These modes are active along both $SiO_2$ layers as well as the top and bottom Ag surfaces. The spatial mapping of the electric field indicates that the modes in the two $SiO_2$ insulators are coupled, because they demonstrate a characteristic beating pattern that indicates that power is being transferred between the two MIMs. This result was expected physically—the spacer layer between the two insulating layers is thinner than the skin depth of Ag at the energies of the generated electric fields.

To better determine the natures of the various modes within the MSPF, a second Fourier Transform was performed. By taking an FFT over the propagation direction of the modes, the phasor direct space dataset can be moved into momentum space (i.e. "k-space"). The results of this FFT can be plotted, as shown in Figure 3(b), to reveal the spatially resolved intensity of the various modes within a structure that was excited by a single frequency source. The profiles of these modes can be determined by spatially mapping the intensity a given spectral frequency. For example, in Figure 3(b) there is one hot spot at 2.7 cycles per micron that spans the MSPF from 60 nm to 300 nm, one hot spot on the top surface of the filter at 1.4 cycles per micron that spans 300 nm to over 400 nm, and a collection of hot spots below the bottom surface of the MSPF (at spectral frequencies spanning 0.5 to 2.3 cycles per micron).

Closer inspection of the intensity patterns of the hot spots below 70 nm shows that those at the lower spectral frequencies are unbound modes that do not contribute to filter behavior. Discarding those hot spots and plotting the rest of the linear intensity variations of the single spectral frequency hot spots yields the plot shown in Figure 3(c). This plot indicates there are three active plasmon modes in this structure, on the top surface of the filter, one on the bottom surface of the filter, and one spanning the insulating layers contained within the filter. The predominant surface mode, shown in blue in Fig 3(c), corresponds to SPPs excited at the top Ag surface of the filter. The other excitation is a super-mode corresponding to a coupling of the two MIM modes generated within each of the two insulating layers in the structure. This is the mode that was implied in the spatial field map in Figure 3(a) is now clearly depicted in Figure 3(c), which reveals that the two MIM modes within the super-mode are coupled because of strong field overlap within the 50 nm Ag spacer.

The behavior of the energy propagating through the filter can be determined by the FFT analysis, but it does not indicate how these modes contribute to the overall behavior of the filter. By normalizing the transmission curves over the dispersion behavior of each mode and the pitches of the filter gratings, we can reveal the exact ways the modes are contributing to the behavior of the unsuppressed transmission peak.

The dispersion behavior of each plasmon mode can be determined by constructing a dispersion curve for the MSPF. The dispersion curve shown in Figure 4(a) was constructed by using the Fourier Transformed k-space data sets and plotting the power of the modes at each spectral frequency as a function of energy. The two branches on curve correspond to the bottom side SPP and the metal-insulator-metal-insulator-metal (MIMIM) super-mode, and can be mapped to the frequency values that correspond to these modes in Figure 3(c). The lower intensity signal plotted to the left of the branches corresponds to the unbound quasi-modes bouncing off the surface of the filter. From the data contained within this plot, the dispersion curve for each of the two modes can be analytically determined.

The transmission behavior was first normalized over the pitch of the filter grating to lift the dependence of the transmission curves on that characteristic of the filters3. Figures 4(b) and S1(b) show the transmission curves normalized by the SPP and MIMIM dispersion curves respectively. The transmission peaks normalized by the SPP dispersion curve are aligned between 0.5 and 1, with the transmission minima collapsing at 1 on the normalized axis[3]. This behavior illustrates that the SPP mode satisfies the momentum matching conduction required for it to contribute to the transmission behavior of the filter, with the MIMIM super-mode acting as a supplementary suppression to remove second highest order peak.

**Experimental Verification**

MSPFs were fabricated by depositing alternating layers of Ag and $SiO_2$ in an electron beam evaporator and then subsequently milled using a focused ion beam (FIB). The 50 nm slit milled into a 330 nm structure is a prohibitively demanding aspect ratio for a FIB trench mill. For a set of proof-of-concept filters, these demanding design conditions can be relaxed by considering filters only towards the lower energy portion of the visible spectrum. For a slit width of 120 nm, the suppression of the spurious transmission peak is retained and the FWHM of the primary transmission peak only takes a 25 nm hit.

When Ag is deposited on $SiO_2$ in an electron beam evaporator, the Ag films grow with a columnar growth mechanism[8]. These films are rough, which increases plasmonic loss, thereby reducing overall transmission intensity of the filter[9]. The roughness of Ag deposited on $SiO_2$ is even more problematic in a multilayer structure like the MSPF because the roughness of each Ag layer compounds. A rough substrate increases the roughness of the film deposited on it due to differences in atomic flux received by areas of the film with positive and negative curvatures that are larger than can be compensated for by surface diffusion[10]. Because the $SiO_2$ conformally deposits on the underlying Ag layer, each Ag layer sees a progressively rougher substrate, leading to a very rough top surface of the MSPF.

By utilizing a seed layer of AgO deposited onto each $SiO_2$ surface, a much smoother Ag film can be deposited[11]. The AgO is deposited by electron beam evaporating Ag in a chamber with an O2 pressure of $9.5 \times 10^{-5}$ torr. Once 2 nm of AgO are on the surface of the $SiO_2$, the deposition is paused and the AgO is held under vacuum. Because AgO is not vacuum stable, the oxygen is pumped out of the film, leaving a thin Ag layer on the surface of the $SiO_2$[11]. The deposition is then resumed and the rest of the Ag is deposited at in a chamber with pressure $2.3 \times 10^{-6}$ torr and no oxygen flow. The roughness of Ag films deposited with this method was measured to have an RMS of 2.56 nm and the top Ag surface of a multilayer deposited with the AgO growth method has an RMS of 2.92 nm.

To further protect the integrity of the filter, a sacrificial layer was put on the top Ag surface. First a thin layer of PMMA was then spun onto the top surface of the MSPF and then another 70 nm layer of Ag was deposited on top of the PMMA. The sacrificial layer protects the top Ag film of the MSPF by confining the worst of the ion beam damage to the surface of the sacrificial layer, rather than the surface of the MSPF. To utilize the best possible resolution of the ion beam, the slits are milled in FEI Versa FIB, at 30 kV and 1.5 pA[12]. The high accelerating voltage and low beam current help compensate for the high aspect ratio of the filter structure. The sacrificial layer is then removed using a heated solvent bath.

The fabricated filters are then measured using a supercontinuum laser with monocrometer set-up that allows for the sample to be illuminated with a narrow bandwidth of incident radiation. A 50X objective takes the collimated light and focuses it down to a 10 um spot size that is shined on the 30 um x 30 um MSPFs. Squares equal in size to the filters were milled 100 um away from each filter and are used to determine the intensity of the laser. All transmitted power was collected by a Si photodiode that was affixed behind the substrate in which the filters and normalization squares have been milled. The experimental response of each filter was determined by normalizing the light transmitted through the filter by that transmitted through its corresponding normalization square. The experimental transmission response generally confirmed the expected narrowband width behavior, as shown in Figure 5(a).

Figure 5(a) shows the experimental transmission response of a prototype filter that has an inter-slit pitch of 475 nm. A cross section of the prototype filter was milled using the FIB. The micrograph of the filter's cross-section, shown in Fig 5(c) reveals that there is a slight taper to the filter structure. Using FDTD simulations, we can compute the transmission behavior of filters with a progressively increasing sidewall taper. The results of these simulations, shown in Figure 5(b) illustrate the importance of the slit sidewalls on the overall behavior of the structure. Using the information gathered from the FDTD simulations, it was determined that to maintain filtering behavior with side lobe suppression, the sidewalls of the slit could not possess greater than a $5^0$ taper. The side lobe in the experimental transmission is due to the $13.7^0$ taper in the fabricated filter.

The FFT analysis indicated that the MSPF super-mode is responsible for the suppression of the spurious transmission peak, and the individual MIM modes are coupled together. As the slits are tapered, the difference between the lengths of the two channels increases, which affects the interference between the two modes, thereby reducing the filtering efficiency of this mode and allowing multiple orders of modes to propagate through the structure.

The polarization response was also experimentally confirmed to match the simulated predications, as shown in Figures 6(a) and (b).

**Conclusions**

A plasmonic color filter with a single narrowband transmission response was designed using FDTD and fabricated to confirm the simulated response. The filter is readily amenable to device integration, with a size well-matched to state of the art CMOS image sensors. The plasmonic filter utilizes a geometry that flexibly allows for precise selection of the spectral bands of interest, allowing for portable electronic devices to be capable of multi- and hyperspectral imaging. The behavior of this filter was analytically determined to arise from a combination of SPP excitations-- the surface SPP mode leads to the enhanced transmission behavior associated with subwavelength plasmonic filters, while the slightly asymmetric MIM super-mode leads to the suppression of the spurious transmission peak that arises in other narrowband plasmonic filter geometries. The MSPF is inherently gated, and this feature will be capitalized on in future work by incorporating transparent conducting oxides into this geometry to create tunable narrowband color filters spanning both the visible and near infrared parts of the spectrum.

**Fabrication Methods**

A fused silica slide was prepared by 5 minute sonication in acetone followed by a rinse with IPA. The alternating metal and insulating layers were all deposited via electron beam deposition in the same chamber to maintain the integrity of the Ag/ $SiO_2$ interfaces. A silver oxide seeding method was used to produce smooth Ag films[11]. 2 nm of Ag is deposited at a rate of 0.1 A/s in a chamber under a pressure of $9.5 \times 10^{-5}$ torr O2. The AgO film is then reduced to an Ag film under vacuum for 10 minutes to yield an Ag film on the surface of the silica substrate. The remaining 68 nm of Ag are deposited at a 0.5 A/s deposition rate followed by a SiO$_2$ deposition deposited at 1.5 A/s under a pressure of 2.3x10$^{-6}$ torr. The remaining Ag and SiO$_2$ layers are deposited using this method. Once the depositions are completed, 90 nm of PMMA is spun onto the Ag surface before depositing another 70 nm of Ag as a sacrificial layer. Ga+ ions at 30 kV and 1.5 pA are used to mill 130 nm wide slits into the MSPF and sacrificial layer stack. Multi-pass milling is used to reduce the taper of the slits—first a rectangle is milled, followed by a frame around the perimeter, to better define the edges and clean off redeposition within the slit. After milling the sacrificial layer is removed using first heated remover PG followed by submerging it in acetone and spraying it with an acetone squirt bottle before rinsing in IPA.

**Acknowledgments**

This work was supported by Sony Corporation, the Hybrid Nanophonics Multidisciplinary Unviersity Research Initiative Grant (Air Force Office of Scientific Research FA9550-12-1-0024), Northrop Grumman Corporation, and the facilities of the Kavli Nanoscience Institute (KNI) at Caltech. LAS acknowledges support from the Resnick Sustainability Institute at Caltech. Helpful discussions with Michelle Sherrott, Max Jones, and Matt Sullivan are also gratefully acknowledged.

**References**

1 Gao, H.; Zhou, W.; Odom, T. W. Advanced Functional Materials 2010, 20, (4), 529-539.
2. Ebbesen, T. W.; Lezec, H. J.; Ghaemi, H.; Thio, T.; Wolff, P. Nature 1998, 391, (6668), 667-669.
3. Pacifici, D.; Lezec, H. J.; Sweatlock, L. A.; Walters, R. J.; Atwater, H. A. Optics Express 2008, 16, (12), 9222-9238.
4. Yokogawa, S.; Burgos, S. P.; Atwater, H. A. Nano Letters 2012, 12, (8), 4349-4354.
5. Zheng, B. Y.; Wang, Y.; Nordlander, P.; Halas, N. J. Advanced Materials 2014, 26, (36), 6318-6323.
6. Dionne, J.; Sweatlock, L.; Atwater, H.; Polman, A. Physical Review B 2006, 73, (3), 035407.
7. Xu, T.; Wu, Y.-K.; Luo, X.; Guo, L. J. Nature Communications 2010, 1, 59.
8. Thompson, C. V. Annual Review of Materials Science 1990, 20, (1), 245-268.
9. Maradudin, A.; Mills, D. Physical Review B 1975, 11, (4), 1392.
10. Mazor, A.; Srolovitz, D.; Hagan, P.; Bukiet; BG. Physical review letters 1988, 60, (5), 424.

11. Martin, O. J.; Thyagarajan, K.; Santschi, C. In A New Fabrication Method for Aluminum and Silver Plasmonic Nanostructures, Workshop on Optical Plasmonic Materials, 2014; Optical Society of America: p OW3D. 2.

12. Joshi-Imre, A.; Bauerdick, S. Journal of Nanotechnology 2014, 2014.

Figure 1

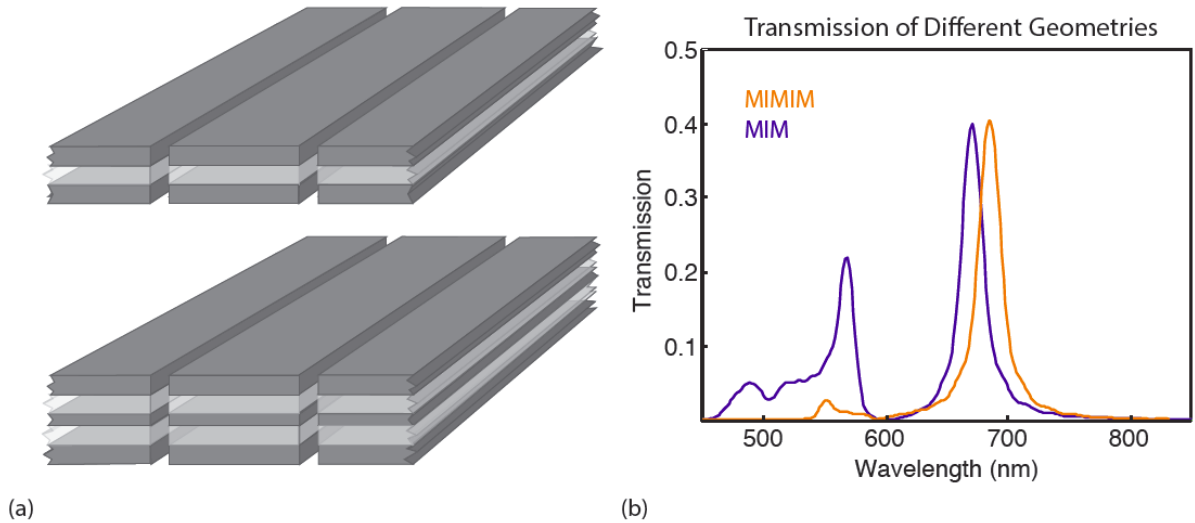

(a) Schematics of MIM and MIMIM filter structures. All dark grey metal layers are Ag and 70 nm thick, except for the 50 nm center metal layer of the MIMIM filter. All light grey insulating layers are 70 nm of SiO2 (b) Comparison between MIM and MIMIM transmission behavior for the structures shown in (a) that shows similar FWHM but enhanced suppression of the secondary peak in the MIMIM case

Figure 2

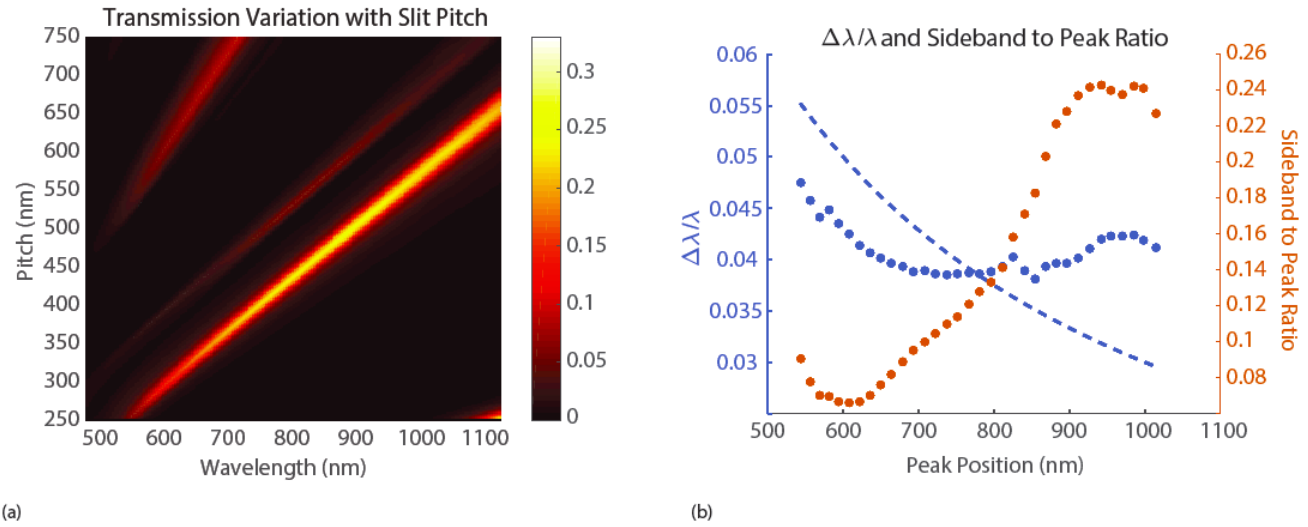

(a) Super position of the transmission behavior of filters with varying slit pitches. As slit pitch increases the narrowband transmission peak is controllably shifted to longer wavelengths (b) The relationship between FWHM, peak position, and sideband to peak ratio. The blue axis illustrates the ratio of FHWM to the peak position. The dashed line sets the threshold of a 30nm FWHM, and the dotted line illustrates the ratio of the transmission peak's FWHM to the peak position. The dotted line is beneath the dashed line for the entire visible spectrum, indicating that all filters fulfill the criteria for hyperspectral imaging. The orange dotted line illustrates the ratio between the sideband and main intensity peaks, showing the best filters are also in the visible part of the spectrum

Figure 3

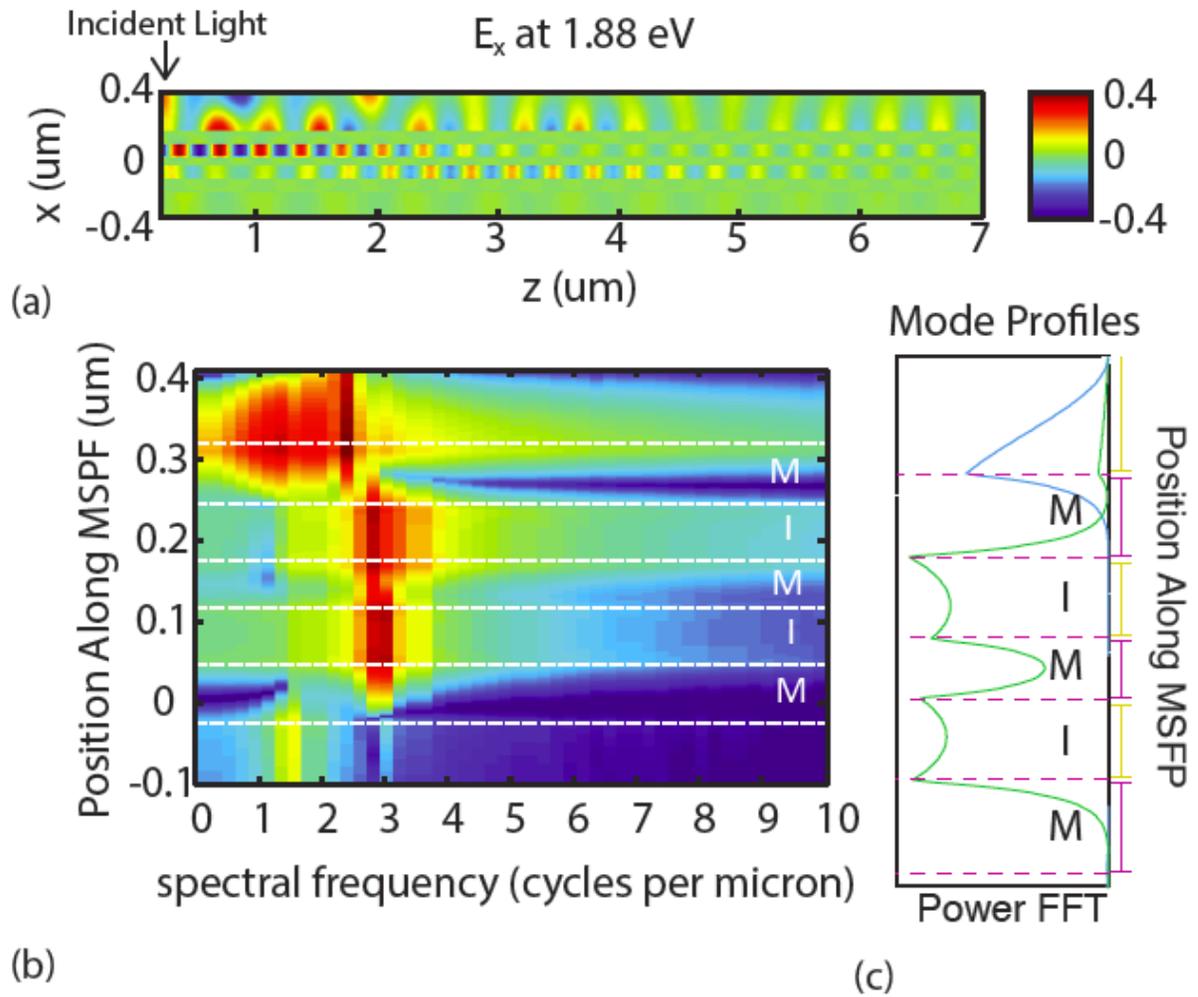

(a) Field behavior from a single slit depicts the coupled interactions of the two MIM modes confined to their insulating layers (b) A FFT of the spatial field behavior of single frequency excitations indicates the presence of multiple modes at different spectral frequencies spanning the structure. (c) These modes can be spatially resolved via a Power FFT

Figure 4

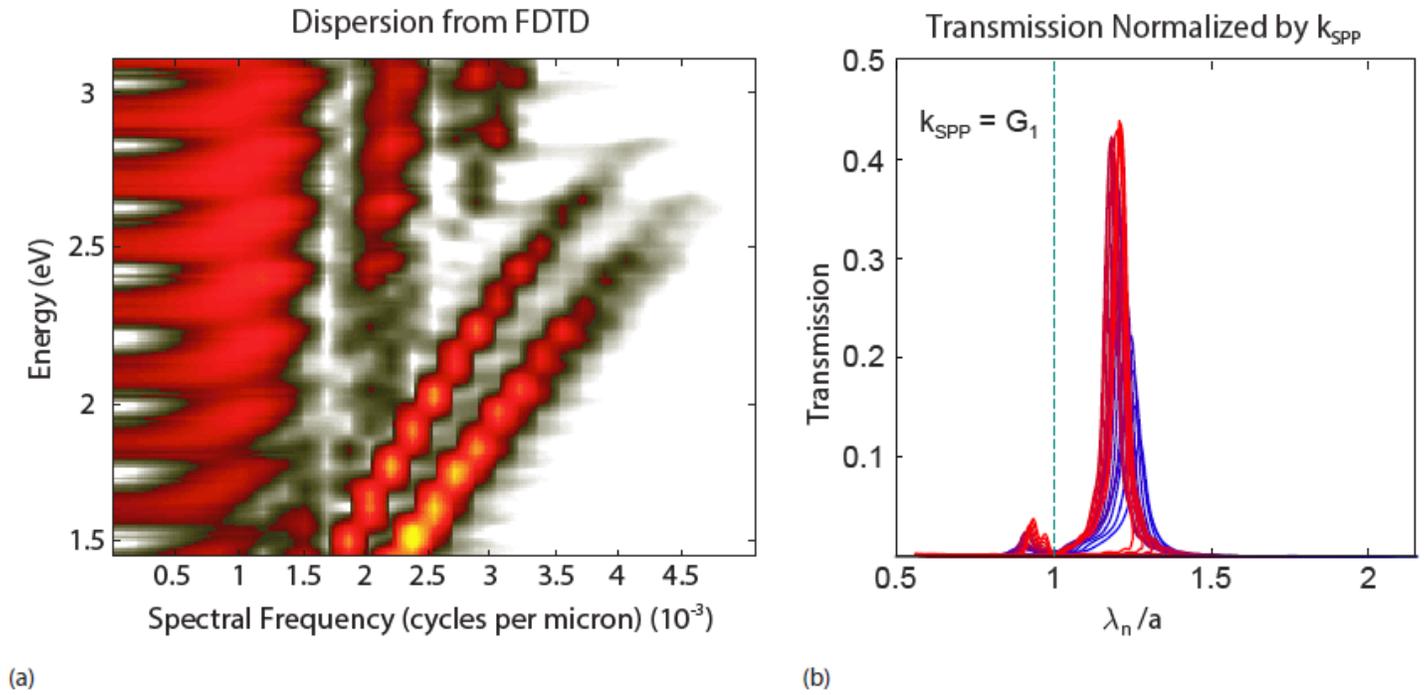

(a)

(b)

(a) By taking FFTs of a sweep of single frequency excitations, a dispersion curve can be constructed that illustrates the behavior of both active modes (b) Universal curve analysis confirms that the SPP mode on the top surface of the MSPF filter is predominantly responsible for the filters transmission behavior. The various colors of the transmission curves correspond to different peak intensity positions that have all been normalized by the SPP dispersion curve.

Figure 5

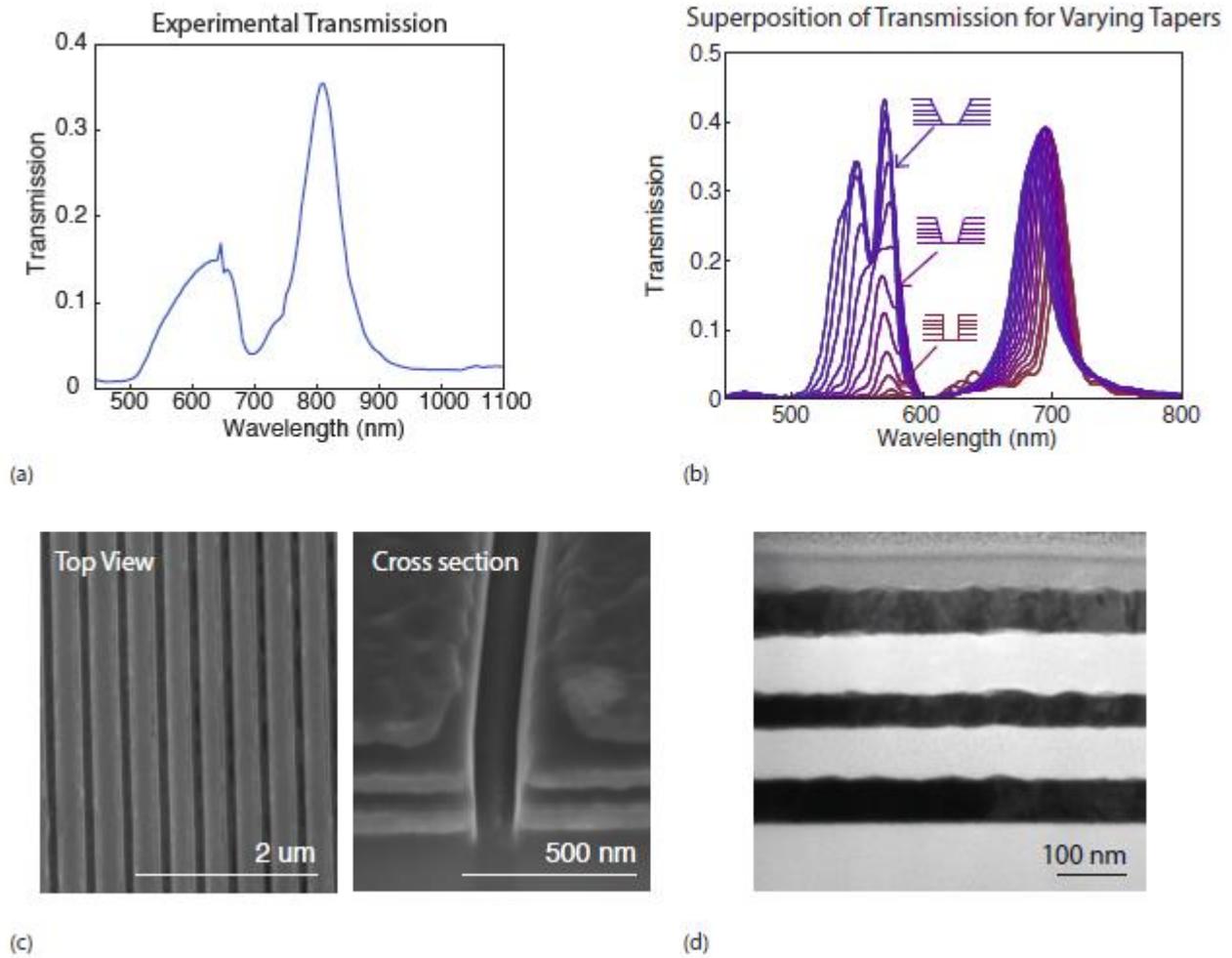

(a) Experimentally determined transmission of a single MIMIM filter (b) Simulated dependence of transmission on taper of slits (c) Top down and cross-sectional SEMs of the MSPF filter (d) TEM micrograph showing the layer thicknesses and roughness of the five layers of the filter

Figure 6

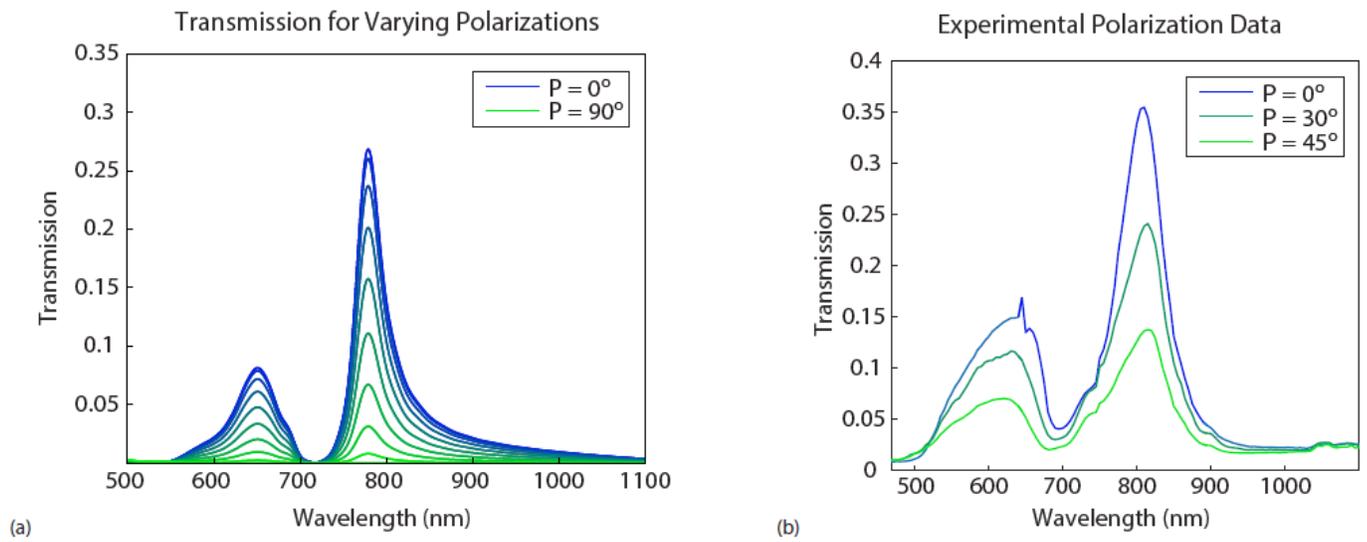

(a) Simulated polarization response varying from 0 degrees (blue) to 90 degrees (green) (b) The simulated polarization response was confirmed experimentally, with a 0 degree measurement (blue), a 30 degree measurement (red), and a 45 degree measurement (orange)

Supplemental Figure

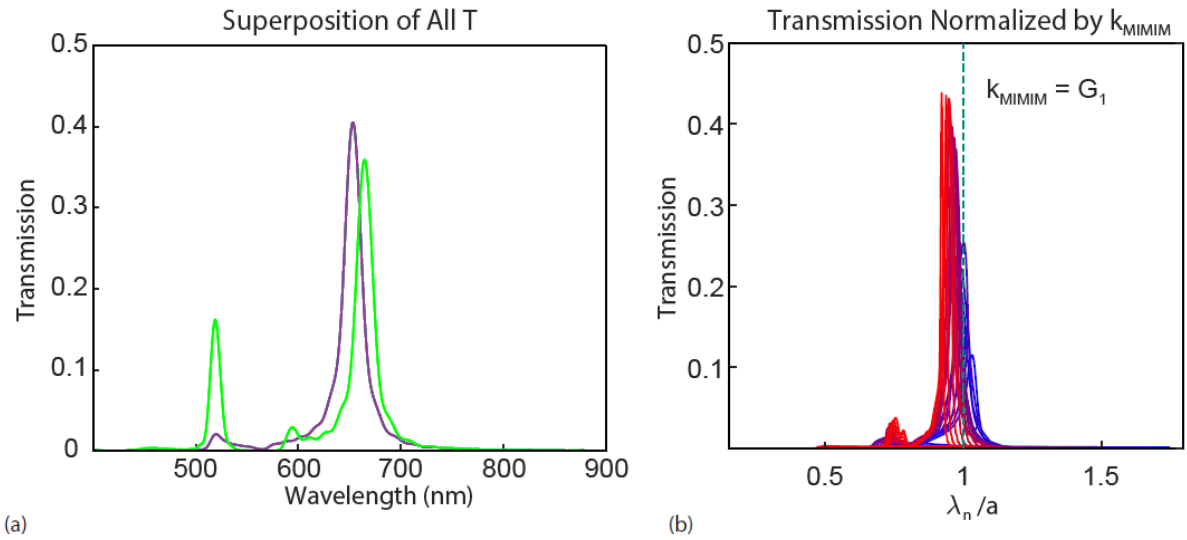

(a)   (b)

(a) A comparison between two MSPF configurations. The violet curve is the asymmetric case, with the top of the MSPF exposed to air and the bottom deposited on an $SiO_2$ substrate. The green curve is the symmetric case, where both the top and bottom of the filter are adjacent to $SiO_2$. In the asymmetric case, the spurious side-lobe transmission is maximally suppressed—in this case the two constituent MIM modes have slightly different indices due to differences in their constituent materials. This difference leads to a slight difference in the amount of accumulated phase, which allows for enhanced destructive interference that is not possible in the symmetric case, leading to a more highly suppressed side lobe. (b) Universal curve analysis illustrates that the MIMIM super-mode is not the dominant mode for transmission. The various colors of the transmission curves correspond to different peak intensity positions that have all been normalized by the SPP dispersion curve.